\documentclass[aps,prl,twocolumn,
groupedaddress,nofootinbib,nobalancelastpage,nobibnotes]{revtex4}
\usepackage{amsmath,amsfonts,amssymb,mathrsfs,graphicx}

\newcommand{\ie}{{\it i.e.}}

\newcommand{\eg}{{\it e.g.}}

\newcommand{\cf}{{\it cf.}}

\newcommand{\fig}{Fig.}

\newcommand{\Ref}{Ref.}

\newcommand{\Tab}{Tab.}

\newcommand{\NuFactII}{\mbox{\sf NuFact-II}}

\newcommand{\stheta}{\sin^22\theta_{13}}
\newcommand{\deltacp}{\delta_\mathrm{CP}}
\newcommand{\ldm}{\Delta m_{31}^2}
\newcommand{\sdm}{\Delta m_{21}^2}

\newcommand{\figu}[1]{\fig~\ref{fig:#1}}

\newcommand{\bi}{\begin{itemize}}
\newcommand{\ei}{\end{itemize}}

\begin{document}

\title{Probing the absolute density of the Earth's core using a neutrino beam}

\author{Walter Winter}
\affiliation{School of Natural Sciences, Institute for Advanced Study, Einstein Drive,
Princeton, NJ 08540}

\date{\today}

\begin{abstract}
\vspace*{0.2cm}
We demonstrate that one could measure the absolute matter density of the Earth's core
with a vertical neutrino factory baseline at the per cent level for $\stheta \gtrsim 0.01$,
where we include all correlations with the oscillation parameters in the analysis. We discuss
the geographical feasibility of such an approach, and illustrate how the results change as
a function of the detector location. We point out the complementarity to geophysics.
\end{abstract}


\maketitle


\begin{figure}[t]
\begin{center}
\includegraphics[width=\columnwidth]{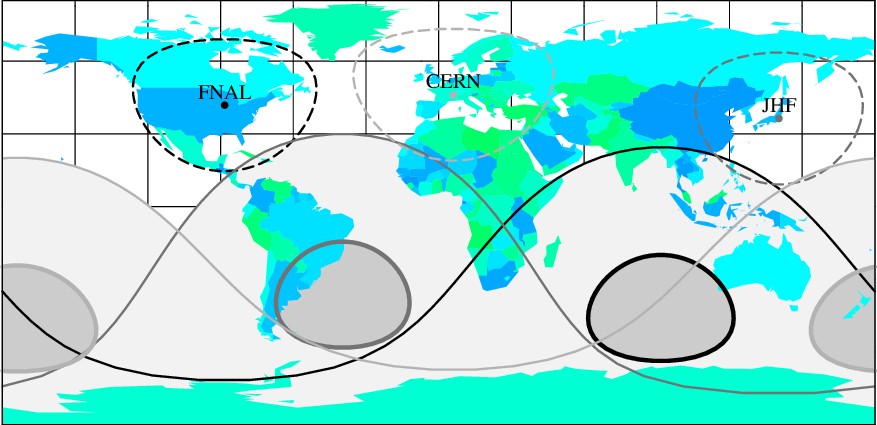}
\end{center}
\caption{\label{fig:labs} (Color online) Positions of three of the major potential neutrino
factory laboratories, (typical) $L=3 \, 000 \, \mathrm{km}$ detector sites (dashed curves),
as well as
potential detector sites with outer core crossing baselines (below thin solid curves), and inner core crossing baselines (within thick solid curves). The colors of the curves
represent the different laboratories.}
\end{figure}

Neutrino oscillation physics has entered the age of precision physics, which means
that the leading atmospheric and solar oscillation parameters are known to high precisions
and the next generation of long-baseline experiments will be highly sensitive to
sub-leading effects. These long-baseline experiments send an artificially produced
neutrino beam of energy $E$ on a straight baseline (length $L$) through the Earth to a detector, which is, depending on the neutrino energy, several hundred to many thousand kilometers away. In particular, the future potential high precision instrument ``neutrino factory''~\cite{Geer:1998iz%
}  leads to typical baselines $\sim 700 \, \mathrm{km} - 7 \, 500 \, \mathrm{km}$ relevant for neutrino oscillation physics.
It is an interesting feature of neutrino oscillations
that the flavor conversion is sensitive to the electron density
$n_e$ of Earth matter~\cite{Wolfenstein:1978ue%
}, which has
been suggested to be used for neutrino oscillation tomography of the Earth's interior~\cite{Ermilova:1988pw,Ohlsson:2001ck,Lindner:2002wm%
}. The electron density then translates into the matter density
by $\rho = n_e \, m_N /Y_e$ with $Y_e = Z/A$ the ``electron fraction'' and $m_N$ the nucleon mass. For a neutrino factory, the signal amplitude of such a measurement using the flavor conversion
$\nu_\mu \leftrightarrow \nu_e$
is given by the parameter $\stheta$, which has so far only been constrained by the
CHOOZ experiment to $\stheta \lesssim 0.1$~\cite{Apollonio:1999ae}. Future long baseline
experiments will find $\stheta>0$ within the coming ten years if $\stheta \gtrsim 0.01$ (see, \eg, \Ref~\cite{Huber:2004ug}).

The most successful approach to the tomography of the Earth's interior has been seismic wave
geophysics primarily using seismic waves from earthquakes to reconstruct a profile of the Earth's interior. Most of the energy produced by an earthquake
is deposited in shear waves (s-waves), which cannot penetrate into the Earth's (outer) liquid core (but might be partially converted into p-waves). A smaller fraction of energy goes into
pressure waves (p-waves), which are propagated into the Earth's core, too. The
waves are partially reflected at the mantle-outer core and outer core-inner core boundaries.
 Therefore, seismic waves are highly sensitive to density jumps and the positions of these boundaries. Seismic waves geophysics leads to a propagation velocity profile of the Earth's interior, which can be translated into a density profile with the equation of state. This conversion is based on a model for the shear/bulk modulus, which implies uncertainties. In fact, the most direct information on the matter density
 distribution comes from the Earth's mass and its moment of inertia about the polar axis, which are,
 however, not uniquely determining it. In particular, there are
many open questions about the Earth's inner core (see, \eg, \Ref~\cite{Steinle-Neumann:2002}). Therefore, a measurement of the absolute electron density
of the Earth's core could provide very complementary information to geophysics.

From the point of view of a neutrino factory, this would require a ``vertical'' baseline,
\ie, a decay tunnel which is vertical. Since there are only a number major high energy laboratories which
are candidates for a neutrino factory, all of these on the northern hemisphere, the
geographical aspect is another important part of this problem. As it is illustrated in
\figu{labs}, there are indeed possible detector locations (\ie, on land instead of in water)
for many of the major laboratories with core crossing baselines.
For example, a baseline from CERN to New Zealand crosses the inner core of the Earth.


\begin{figure}
\begin{center}
\includegraphics[width=0.9\columnwidth]{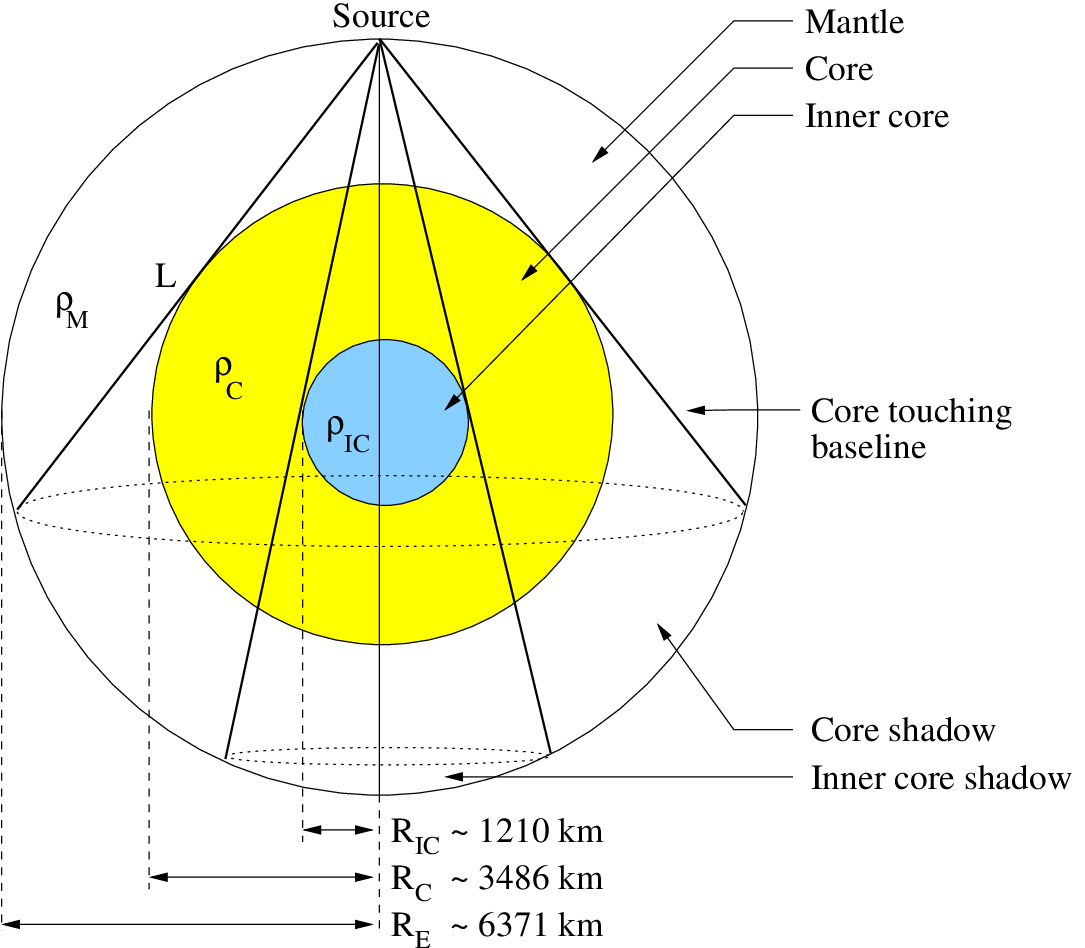}
\end{center}
\caption{\label{fig:earthpic} (Color online) The
geometry of the core tomography problem (not to scale).}
\end{figure}

The geometry and the quantities of interest of the Earth tomography problem are illustrated in \figu{earthpic}.
Though neutrino oscillations are, in principle, sensitive to
structural features of the density profile,
realistic experiments with only one baseline have only
very limited abilities for a detailed reconstruction because
they are insensitive to structures much shorter than the
oscillation length in matter~\cite{Ohlsson:2001ck}.
Therefore, it is important that the relevant independent
parameters be identified. In particular, treating the
(average) outer and inner core densities $\rho_C$ and $\rho_{IC}$ as
independent parameters (\cf, \figu{earthpic}) implies strong correlations~\cite{Lindner:2002wm}.
However, these two quantities are not really independent, since
the total mass of the Earth is extremely well known. In addition,
we have very good knowledge on the Earth's mantle from seismic
wave geophysics.
If we furthermore assume that the positions of the
mantle-outer core and outer core-inner core boundaries are very well known from
the reflections of seismic waves, we can reduce the number of parameters
to one (either $\rho_C$ or $\rho_{IC}$). Below, we will argue that it is always
reasonable to measure the average density of the ``innermost'' shell
a baselines crosses.

Neutrino oscillations are, for the slowly enough varying matter density within each shell
 $\rho_i$ (mantle, outer core, or inner core), to a first approximation determined by the baseline averaged density
$\bar{\rho}^L_i = 1/L \int_0^{L} \tilde{\rho}(l) dl $~\cite{Ohlsson:2001et}. Here $\tilde{\rho}(l)$ is the density along the baseline.
 On the other hand, the average density of
the Earth is determined by the volume averaged density, which is, for each shell, given by
$\bar{\rho}^V_i = 4 \pi/V \int_{R_1}^{R_2} \rho(r) r^2 dr$.
Here $r$ is the distance from the Earth's center and $V = 4/3 (R_2^3 - R_1^3) \pi$ the volume
of the shell. This means that the density within each differential shell $dr$ is weighted with $r^2$. For the total mean density of the Earth we then have
\begin{equation}
\bar{\rho}^V = (\bar{\rho}^V_{IC} - \bar{\rho}^V_C) \left( \frac{R_{IC}}{R_E} \right)^3 + (\bar{\rho}^V_{C} - \bar{\rho}^V_M) \left( \frac{R_{C}}{R_E} \right)^3 + \bar{\rho}^V_M
\nonumber
\end{equation}
The constraint $M_E = const$, \ie,
$\bar{\rho}^V=const$ because of the (approximately) known volume, together with the
assumption of a known/fixed
$\bar{\rho}^V_M$, leads by differentiation with respect to $\bar{\rho}^V_{IC}$ to
$
\Delta \bar{\rho}^V_C = - \left( R_{IC}/R_C \right)^3 \, \Delta \bar{\rho}^V_{IC} \simeq - 0.04 \, \Delta \bar{\rho}_{IC}$.
Thus, a large change in the measurement in $\bar{\rho}^V_{IC}$ can be compensated by a very
small change in $\bar{\rho}^V_C$ because of the volume averaging.
The same is, in principle, true for the outer core and mantle densities. Since the effect on
neutrino oscillations is proportional to the baseline length within each shell and not to
the volume, each shell contributes to the total averaged density $\bar{\rho}^L$ by a similar magnitude. It is therefore reasonable to treat the density of the innermost shell
as baseline crosses as the parameter of interest and correct the (better known) density of the next outer shell by $\Delta \bar{\rho}^V$.
One advantage of the sensitivity to $\bar{\rho}^L$ instead of $\bar{\rho}^V$
is that the actual positions of the boundaries between the different shells only enter
as a second order effect. For example, a shift $R_{IC} \rightarrow R_{IC} + \Delta R$
leads, for a vertical baseline $L=2 \cdot R_E$, to a correction of $\Delta \bar{\rho}^L_{IC} \simeq 2 \frac{\Delta R}{L} (\bar{\rho}_{IC} - \bar{\rho}_{C})$. This means that the positions of the boundaries have to be known to about $\sim 50 \, \mathrm{km}$
for a percent level density measurement.


Based on the above discussion, use $\bar{\rho}^L_{IC} \equiv \bar{\rho}_{IC}$ as parameter for inner core crossing baselines, and $\bar{\rho}^L_C \equiv \bar{\rho}_C$ for baselines which only cross the outer core (\cf, \figu{earthpic}). In fact, because of the actual sensitivity to the electron density, we
measure the effective density $\rho^{\mathrm{eff}} = \sum_i \rho_i \, Y_e^i / 0.5$ averaged over the baseline, where the index $i$ refers to the contained elements (for example, for iron $Y_e \simeq 0.464$).
Since for most of the elements in the core $Y_e$ is close to $0.5$ and therefore very similar, errors in the composition translate into errors in $\bar{\rho}$ only as second order effect.
We assume the matter densities to be constant within each shell, \ie, we ignore the
matter profile effect, which corresponds to measuring the absolute normalization of the Reference
Earth Model (REM)~\cite{Dziewonski} profile. We use a complete simulation of the neutrino factory \NuFactII\ from \Ref~\cite{Huber:2002mx} simulated with the GLoBES software~\cite{Huber:2004ka}.
In particular, we include statistics, systematics, and
(connected and disconnected) degenerate solutions
with the oscillation parameters, \ie, we marginalize with respect to the oscillation
parameters. This procedure is necessary to test if the matter effect sensitivity survives a realistic simulation taking into account the insufficient knowledge on the oscillation parameters.
The neutrino factory uses muons with an energy of $50 \, \mathrm{GeV}$,
$4 \, \mathrm{MW}$ target power ($5.3 \cdot 10^{20}$
useful muon decays per year), and a magnetized iron detector with a fiducial
mass of $50 \, \mathrm{kt}$. We choose a symmetric operation with $4 \, \mathrm{yr}$ in
each polarity. For the oscillation parameters, we use
the current best-fit values $\ldm = 2.5 \cdot 10^{-3} \, \mathrm{eV}^2$,
$\sin^2 2 \theta_{23} = 1$, $\sdm = 8.2 \cdot 10^{-5} \, \mathrm{eV}^2$, and $\sin^2 2 \theta_{12} = 0.83$~\cite{Fogli:2003th%
}.  We only allow values for $\stheta$ below
the CHOOZ bound $\stheta \lesssim 0.1$~\cite{Apollonio:1999ae} and choose $\deltacp=0$
as well as a normal mass hierarchy, where the results should hardly depend on the choices of the latter two parameters. Furthermore, for the leading solar parameters, we impose
external precisions of $10\%$ on each $\sdm$ and $\theta_{12}$~\cite{Gonzalez-Garcia:2001zy}.

\begin{figure}[t]
\begin{center}
\includegraphics[width=0.9\columnwidth]{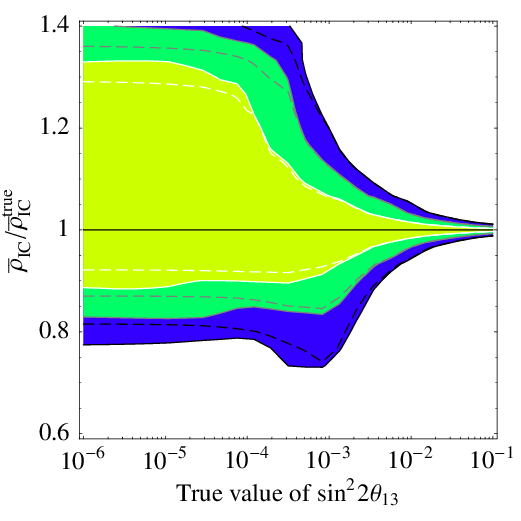}
\end{center}
\caption{\label{fig:th13dep} (Color online) The measurement of $\bar{\rho}_{IC}$ as function of the true value of $\stheta$ at the $1\sigma$, $2\sigma$, and $3 \sigma$ confidence levels (from light to dark shaded regions).
For the baselines, $L=2 \cdot R_E$ combined with $L=3 \, 000 \, \mathrm{km}$ is used. The
dashed curves correspond to fixing the oscillation parameters, \ie, to not taking into
account correlations and degeneracies.
}
\end{figure}


\begin{table}[b]
\begin{center}
\begin{tabular}{lcccc}
\hline
& \multicolumn{2}{c}{$L=2 \cdot R_E$}  &
\multicolumn{2}{c}{$L=12 \, 510 \, \mathrm{km}$}  \\
 & \multicolumn{2}{c}{\% error on $\bar{\rho}_{IC}$} &
\multicolumn{2}{c}{\% error on $\bar{\rho}_{C}$} \\
$\stheta$ & $1 \sigma$ & $3 \sigma$ & $1 \sigma$ & $3 \sigma$ \\
\hline
\multicolumn{5}{c}{{\bf Combination with $\boldsymbol{L=3 \, 000 \, \mathrm{km}}$:}} \\[0.1cm]
$0.1$ & $-0.5$/$+0.5$ & $-1.4$/$+1.4$ & $-0.2$/$+0.2$ & $-0.6$/$+0.6$ \\[0.1cm]
$0.01$ & $-1.8$/$+1.7$ & $-5.5$/$+5.0$ & $-0.6$/$+0.6$ & $-1.8$/$+1.7$ \\[0.1cm]
$0.001$ & $-8.3$/$+6.9$ & ${-27}$/${+21}$ & ${-1.8}$/${+2.2}$ & ${-4.9}$/${+7.0}$ \\[0.2cm]
\multicolumn{5}{c}{{\bf Core crossing baseline alone:}} \\[0.1cm]
$0.1$ & ${-0.5}$/${+0.5}$ & ${-1.4}$/${+1.4}$ & ${-0.3}$/${+0.2}$ & ${-0.8}$/${+0.6}$ \\[0.1cm]
$0.01$ & ${-2.1}$/${+5.8}$ & ${-7.2}$/${+9.2}$ & ${-0.9}$/${+0.9}$ & ${-2.4}$/${+2.7}$ \\[0.1cm]
$0.001$ & ${-9.9}$/${+19}$ & ${-40}$/${+35}$ & ${-2.3}$/${+2.5}$ & ${-14}$/${+10}$ \\[0.1cm]
\hline
\end{tabular}
\end{center}
\caption{\label{tab:results} Per cent errors ($1 \sigma$ and $3 \sigma$) on $\bar{\rho}_{IC}$ for
 an inner core crossing baseline and $\bar{\rho}_{C}$ for an inner core touching baseline as labeled in the columns.
The upper group of numbers refers to the combination with a baseline $L=3 \, 000 \, \mathrm{km}$,
the lower group to the core crossing baseline alone. The numbers are given for different values of $\stheta$ and $\deltacp=0$.
}
\end{table}

\begin{figure*}
\begin{center}
\includegraphics[width=0.32\textwidth]{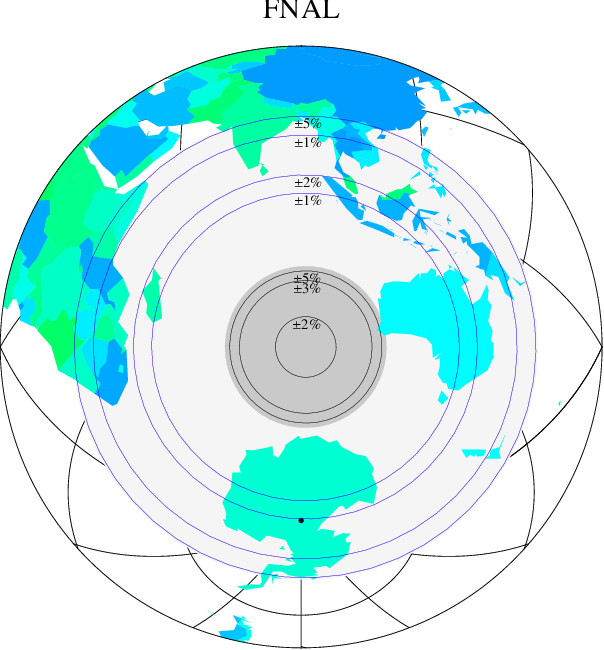} \hspace{0.1cm} %
\includegraphics[width=0.32\textwidth]{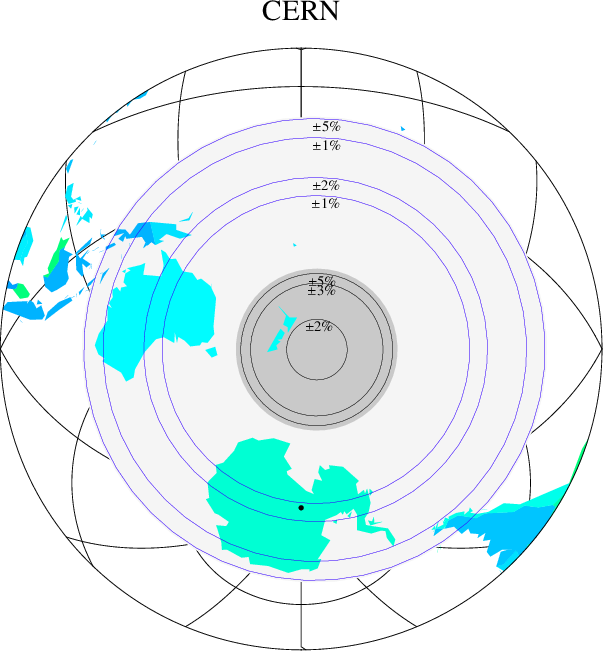} \hspace{0.1cm} %
\includegraphics[width=0.32\textwidth]{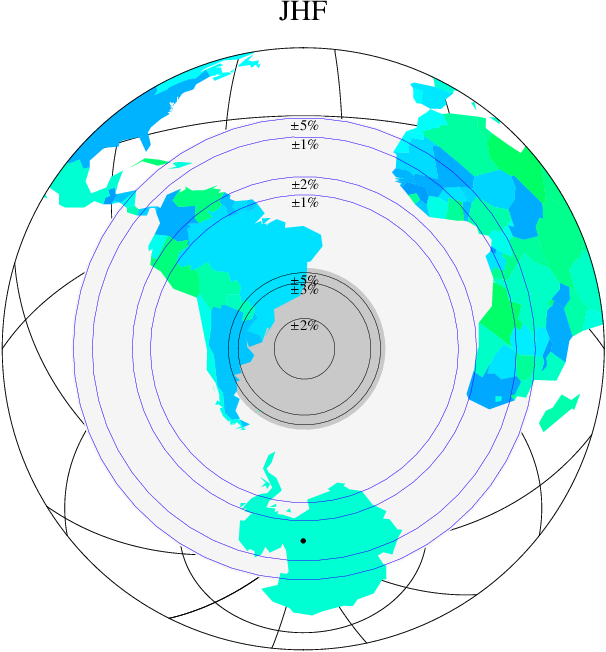}
\end{center}
\vspace*{-0.3cm}
\caption{\label{fig:siteres} (Color online) Precision of the measurement of $\bar{\rho}^L_C$
(light gray shaded outer core crossing regions) or $\bar{\rho}^L_{IC}$
(dark gray shaded inner core crossing regions) as function of the actual position of the detector
for the three laboratories in \figu{labs} as neutrino sources (from west to east).
The $1 \sigma$ errors are given within the rings as they are labeled.
For all setups, the
combination with $L= 3 \,000 \, \mathrm{km}$, as well as $\stheta=0.01$ are assumed.}
\end{figure*}

In principle, it is enough to have one baseline to measure the average density of the Earth's
core. However, if a neutrino factory is built, it's main purpose will be the determination
of $\deltacp$ and other sub-leading effects, which means that (for a muon energy of $50 \, \mathrm{GeV}$)
$L= 3 \, 000 \, \mathrm{km}$ is a typical baseline for that purpose (\cf, \figu{labs}
for potential detector sites). Therefore, we show in
\figu{th13dep} the precision of the measurement of $\bar{\rho}_{IC}$ with $L=2 \cdot R_E$
as function of the critical parameter $\stheta$ in combination with $L= 3 \, 000 \, \mathrm{km}$ to reduce the errors of the oscillation parameters.
Note that we have chosen the longest possible core-crossing baseline, which is a rather optimistic assumption at first.
\figu{th13dep} demonstrates that $\stheta \gtrsim 10^{-3}$ is a prerequisite to obtain high precisions, which also applies to the measurement of $\bar{\rho}_{C}$
(not shown). By comparison of the shaded contours (includes correlations) and the dashed curves (no correlations),
it also demonstrates that the combination with $L= 3 \, 000 \, \mathrm{km}$ is very close to the optimal (correlation-free) performance. In \Tab~\ref{tab:results}, the per cent errors are listed for different selected values of
$\stheta$, where also the values for the core crossing baseline alone and for the
measurement of $\bar{\rho}_{C}$ for an inner core touching baseline are shown. From
\figu{th13dep} and \Tab~\ref{tab:results} one can easily see that each core crossing baseline
alone is almost correlation-free for large values of $\stheta$, whereas for small values
of $\stheta$ the $L=3 \, 000 \, \mathrm{km}$ baseline provides valuable additional
information to reduce correlations. The reason for this is that the correlation with $\deltacp$
is rather unimportant for large values of $\stheta$ at these very long baselines, where
the contribution from the solar oscillations becomes MSW effect suppressed (see, \eg, \Ref~\cite{Huber:2003ak}). From \Tab~\ref{tab:results}, we can finally
read off a $1 \sigma$ error of less than one per cent for $\rho_{IC}$ under optimal conditions, \ie, for large $\stheta$ and long enough baselines.


Let us now restrict these optimal assumptions somewhat. As it is obvious from \figu{labs},
there may not always be potential detector locations especially for inner core crossing baselines.
Therefore, we show in \figu{siteres} the precision of the measurement for a somewhat
smaller value of $\stheta = 0.01$ as function of the actual detector location for
the laboratories as given in the plot labels. In this case, the optimal precision for the
inner core crossing baselines of about $2\%$ can not be reached because there is
nothing besides water on the opposite sides of the main laboratories. However, a precision
of $3\%$ is easily reachable, since, because of the spherical geometry, baselines through the
inner core travel long distances within the inner core for a large region of the inner core shadow region projected onto the Earth's surface. The same, in principle, applies to the outer core crossing baselines.


In summary, a precision determination of the Earth's absolute inner or outer core density at
the percent level seems to be feasible with a neutrino factory baseline for large enough
$\stheta \gtrsim 10^{-2}$, even if one takes into account the knowledge on the neutrino oscillation parameters. Since a neutrino factory muon decay ring naturally spans two baselines, such a core crossing baseline could be an additional
(or subsequent) payoff of a neutrino factory in addition to its main purpose to measure the
neutrino oscillation parameters precicely. The major challenge of such an approach would be building am appropriate decay ring with a vertical decay tunnel. From the geographical point of view, given some of the current potential candidates for a neutrino factory, there are many
potential detector locations on the southern hemisphere where sufficient precisions could be obtained. From the neutrino physics point of view, other applications of such a very
long baseline would be the degeneracy-free measurement of $\stheta$
or the verification of the MSW effect~\cite{Huber:2003ak}.

In comparison to seismic wave geophysics, neutrinos are sensitive to the absolute electron
density in matter, which means that the relationship to the absolute matter density is
much cleaner from model-dependent assumptions. This measurement could therefore help
to test the equation of state for seismic waves. The obtainable relative precision at the
percent level is competitive to the relative precisions of the density jumps given by seismic wave geophysics. For example, at the inner core boundary, $\Delta \rho \simeq 0.82 \pm 0.18 \, \mathrm{g} \, \mathrm{cm}^{-3}$~\cite{Masters}.
Of course, the neutrino factory approach alone is unlikely to give a three-dimensional model
of the Earth, but its strength to measure the baseline averaged density as opposed to the
volume averaged density makes it a good candidate for
direct matter density tests of especially the innermost parts of the Earth.


I would like to thank Peter Goldreich and Jeroen Tromp for useful information and discussions, and Pomita Ghoshal for pointing out an error. This work has been supported by the W.~M.~Keck foundation.

\vspace*{-0.2cm}


{\small

\begin{thebibliography}{10}
\expandafter\ifx\csname bibnamefont\endcsname\relax
  \def\bibnamefont#1{#1}\fi
\expandafter\ifx\csname bibfnamefont\endcsname\relax
  \def\bibfnamefont#1{#1}\fi
\expandafter\ifx\csname url\endcsname\relax
  \def\url#1{\texttt{#1}}\fi
\expandafter\ifx\csname urlprefix\endcsname\relax\def\urlprefix{URL }\fi
\providecommand{\bibinfo}[2]{#2}
\providecommand{\eprint}[2][]{\url{#2}}

\bibitem{Geer:1998iz}
\bibinfo{author}{\bibfnamefont{S.}~\bibnamefont{Geer}}, \bibinfo{journal}{Phys.
  Rev.} \textbf{\bibinfo{volume}{D57}}, \bibinfo{pages}{6989}
  (\bibinfo{year}{1998});
\bibinfo{author}{\bibfnamefont{M.}~\bibnamefont{Apollonio}} \emph{et~al.}
  \eprint{hep-ph/0210192};
\bibinfo{author}{\bibfnamefont{C.}~\bibnamefont{Albright}} \emph{et~al.}
  \eprint{physics/0411123}.

\bibitem{Wolfenstein:1978ue}
\bibinfo{author}{\bibfnamefont{L.}~\bibnamefont{Wolfenstein}},
  \bibinfo{journal}{Phys. Rev.} \textbf{\bibinfo{volume}{D17}},
  \bibinfo{pages}{2369} (\bibinfo{year}{1978});
\bibinfo{author}{\bibfnamefont{S.~P.} \bibnamefont{Mikheev}} \bibnamefont{and}
  \bibinfo{author}{\bibfnamefont{A.~Y.} \bibnamefont{Smirnov}},
  \bibinfo{journal}{Sov. J. Nucl. Phys.} \textbf{\bibinfo{volume}{42}},
  \bibinfo{pages}{913} (\bibinfo{year}{1985}).

\bibitem{Ermilova:1988pw}
\bibinfo{author}{\bibfnamefont{V.~K.} \bibnamefont{Ermilova}},
  \bibinfo{author}{\bibfnamefont{V.~A.} \bibnamefont{Tsarev}},
  \bibnamefont{and} \bibinfo{author}{\bibfnamefont{V.~A.}
  \bibnamefont{Chechin}}, \bibinfo{journal}{Bull. Lebedev Phys. Inst.}
  \textbf{\bibinfo{volume}{NO.3}}, \bibinfo{pages}{51} (\bibinfo{year}{1988});
\bibinfo{author}{\bibfnamefont{T.}~\bibnamefont{Ohlsson}} \bibnamefont{and}
  \bibinfo{author}{\bibfnamefont{W.}~\bibnamefont{Winter}},
  \bibinfo{journal}{Europhys. Lett.} \textbf{\bibinfo{volume}{60}},
  \bibinfo{pages}{34} (\bibinfo{year}{2002});
\bibinfo{author}{\bibfnamefont{A.~N.} \bibnamefont{Ioannisian}}
  \bibnamefont{and} \bibinfo{author}{\bibfnamefont{A.~Y.}
  \bibnamefont{Smirnov}} \eprint{hep-ph/0201012};
\bibinfo{author}{\bibfnamefont{A.~N.} \bibnamefont{Ioannisian}}
  \bibnamefont{and} \bibinfo{author}{\bibfnamefont{A.~Y.}
  \bibnamefont{Smirnov}}, \bibinfo{journal}{Phys. Rev. Lett.}
  \textbf{\bibinfo{volume}{93}}, \bibinfo{pages}{241801}
  (\bibinfo{year}{2004});
\bibinfo{author}{\bibfnamefont{A.~N.} \bibnamefont{Ioannisian}} \emph{et~al.}
  \eprint{hep-ph/0407138}.

\bibitem{Ohlsson:2001ck}
\bibinfo{author}{\bibfnamefont{T.}~\bibnamefont{Ohlsson}} \bibnamefont{and}
  \bibinfo{author}{\bibfnamefont{W.}~\bibnamefont{Winter}},
  \bibinfo{journal}{Phys. Lett.} \textbf{\bibinfo{volume}{B512}},
  \bibinfo{pages}{357} (\bibinfo{year}{2001}).

\bibitem{Lindner:2002wm}
\bibinfo{author}{\bibfnamefont{M.}~\bibnamefont{Lindner}} \emph{et~al.},
  \bibinfo{journal}{Astropart. Phys.} \textbf{\bibinfo{volume}{19}},
  \bibinfo{pages}{755} (\bibinfo{year}{2003}).

\bibitem{Apollonio:1999ae}
\bibinfo{author}{\bibfnamefont{M.}~\bibnamefont{Apollonio}} \emph{et~al.}
  (\bibinfo{collaboration}{CHOOZ}), \bibinfo{journal}{Phys. Lett.}
  \textbf{\bibinfo{volume}{B466}}, \bibinfo{pages}{415} (\bibinfo{year}{1999}).

\bibitem{Huber:2004ug}
\bibinfo{author}{\bibfnamefont{P.}~\bibnamefont{Huber}} \emph{et~al.},
  \bibinfo{journal}{Phys. Rev.} \textbf{\bibinfo{volume}{D70}},
  \bibinfo{pages}{073014} (\bibinfo{year}{2004}).

\bibitem{Steinle-Neumann:2002}
\bibinfo{author}{\bibfnamefont{V.}~\bibnamefont{Dehant}} \emph{et~al.}
  \bibnamefont{(editors)}
  \emph{Earth's Core: Dynamics, Structure, Rotation}
  (\bibnamefont{AGU Geodynamic Series, 2003}).

\bibitem{Ohlsson:2001et}
\bibinfo{author}{\bibfnamefont{T.}~\bibnamefont{Ohlsson}} \bibnamefont{and}
  \bibinfo{author}{\bibfnamefont{H.}~\bibnamefont{Snellman}},
  \bibinfo{journal}{Eur. Phys. J.} \textbf{\bibinfo{volume}{C20}},
  \bibinfo{pages}{507} (\bibinfo{year}{2001}).

\bibitem{Dziewonski}
\bibinfo{author}{\bibfnamefont{A. M.}~\bibnamefont{Dziewonski}} \bibnamefont{and}
  \bibinfo{author}{\bibfnamefont{D. L.}~\bibnamefont{Anderson}},
  \bibinfo{journal}{Phys. Earth Planet Int.} \textbf{\bibinfo{volume}{25}},
  \bibinfo{pages}{297} (\bibinfo{year}{1981}).

\bibitem{Huber:2002mx}
\bibinfo{author}{\bibfnamefont{P.}~\bibnamefont{Huber}},
  \bibinfo{author}{\bibfnamefont{M.}~\bibnamefont{Lindner}}, \bibnamefont{and}
  \bibinfo{author}{\bibfnamefont{W.}~\bibnamefont{Winter}},
  \bibinfo{journal}{Nucl. Phys.} \textbf{\bibinfo{volume}{B645}},
  \bibinfo{pages}{3} (\bibinfo{year}{2002}).

\bibitem{Huber:2004ka}
\bibinfo{author}{\bibfnamefont{P.}~\bibnamefont{Huber}},
  \bibinfo{author}{\bibfnamefont{M.}~\bibnamefont{Lindner}}, \bibnamefont{and}
  \bibinfo{author}{\bibfnamefont{W.}~\bibnamefont{Winter}},
  \bibinfo{journal}{Comp. Phys. Comm.}  (\bibinfo{year}{to be published}),
  \eprint{hep-ph/0407333}, \urlprefix\url{http://www.ph.tum.de/~globes}.

\bibitem{Fogli:2003th}
\bibinfo{author}{\bibfnamefont{G.~L.} \bibnamefont{Fogli}} \emph{et~al.},
  \bibinfo{journal}{Phys. Rev.} \textbf{\bibinfo{volume}{D67}},
  \bibinfo{pages}{093006} (\bibinfo{year}{2003});
\bibinfo{author}{\bibfnamefont{J.~N.} \bibnamefont{Bahcall}},
  \bibinfo{author}{\bibfnamefont{M.~C.} \bibnamefont{Gonzalez-Garcia}},
  \bibnamefont{and}
  \bibinfo{author}{\bibfnamefont{C.}~\bibnamefont{Pena-Garay}},
  \bibinfo{journal}{JHEP} \textbf{\bibinfo{volume}{08}}, \bibinfo{pages}{016}
  (\bibinfo{year}{2004});
\bibinfo{author}{\bibfnamefont{A.}~\bibnamefont{Bandyopadhyay}} \emph{et~al.}
  \eprint{hep-ph/0406328};
\bibinfo{author}{\bibfnamefont{M.}~\bibnamefont{Maltoni}} \emph{et~al.},
\bibinfo{journal}{New J. Phys.}
  \textbf{\bibinfo{volume}{6}}, \bibinfo{pages}{122} (\bibinfo{year}{2004}).

\bibitem{Gonzalez-Garcia:2001zy}
\bibinfo{author}{\bibfnamefont{M.~C.} \bibnamefont{Gonzalez-Garcia}}
  \bibnamefont{and}
  \bibinfo{author}{\bibfnamefont{C.}~\bibnamefont{Pe$\tilde{\mathrm{n}}$a-Gara%
y}}, \bibinfo{journal}{Phys. Lett.} \textbf{\bibinfo{volume}{B527}},
  \bibinfo{pages}{199} (\bibinfo{year}{2002}).

\bibitem{Huber:2003ak}
\bibinfo{author}{\bibfnamefont{P.}~\bibnamefont{Huber}} \bibnamefont{and}
  \bibinfo{author}{\bibfnamefont{W.}~\bibnamefont{Winter}},
  \bibinfo{journal}{Phys. Rev.} \textbf{\bibinfo{volume}{D68}},
  \bibinfo{pages}{037301} (\bibinfo{year}{2003}),
\bibinfo{author}{\bibfnamefont{W.}~\bibnamefont{Winter}}
  \eprint{hep-ph/0411309}.

\bibitem{Masters}
\bibinfo{author}{\bibfnamefont{G.}~\bibnamefont{Masters}} \bibnamefont{and}
  \bibinfo{author}{\bibfnamefont{D.}~\bibnamefont{Gubbins}},
  \bibinfo{journal}{Phys. Earth Planet Int.} \textbf{\bibinfo{volume}{140}},
  \bibinfo{pages}{159} (\bibinfo{year}{2003}).

\end{thebibliography}

}


\end{document}